\documentclass{article}%
\usepackage{amsmath}
\usepackage{amsfonts}
\usepackage{amssymb}
\usepackage{graphicx}%
\providecommand{\U}[1]{\protect\rule{.1in}{.1in}}

\begin{document}

\title{Entropic Quantization of Scalar Fields\thanks{Presented at MaxEnt 2014, the
34th International Workshop on Bayesian Inference and Maximum Entropy Methods
in Science and Engineering (September 21--26, 2014, Amboise, France). }}
\author{Selman Ipek, Ariel Caticha\\{\small Physics Department, University at Albany-SUNY, Albany, NY 12222, USA.}}
\date{}
\maketitle

\begin{abstract}
Entropic Dynamics is an information-based framework that seeks to derive the
laws of physics as an application of the methods of entropic inference. The
dynamics is derived by maximizing an entropy subject to constraints that
represent the physically relevant information that the motion is continuous
and non-dissipative. Here we focus on the quantum theory of scalar fields. We
provide an entropic derivation of Hamiltonian dynamics and using concepts from
information geometry derive the standard quantum field theory in the
Schr\"{o}dinger representation.

\end{abstract}

\section{Introduction}

In the entropic dynamics (ED) framework quantum theory is derived as an
application of entropic methods of inference. The framework has been applied
to non-relativistic particles \cite{Caticha 2010} and to relativistic scalar
fields \cite{Caticha 2012} leading to several new insights on foundational
issues such as the nature of time and the problem of measurement in quantum
mechanics. For example, just as in other forms of dynamics, time is defined so
that motion looks simple. In the ED of fields entropic time is defined so that
the fields undergo equal fluctuations in equal times---the field fluctuations
play the role of a clock. The spatial uniformity of these fluctuations
guarantees that time flows at the same rate everywhere. Another appealing
insight is the new light entropic methods cast on the interpretation of the
infinities typical of quantum field theory. We find that the divergences are
not physical but epistemic effects; they are indications that the information
that is relevant for the prediction of certain quantities is very incomplete.

In recent work the ED framework has been substantially improved in several
respects. The early formulations involved assumptions that seemed ad hoc.
Their justification was purely pragmatic --- they worked; they led to the
right answers. For example, use was made of auxiliary variables the physical
interpretation of which remained obscure, and there were further assumptions
about the configuration space metric and the form of the quantum potential. In
\cite{Caticha 2014} it was shown that the auxiliary variables were in fact
unnecessary and could be eliminated. More recently, in \cite{Caticha et al
2014b}, we have shown that ED can lead to a Hamiltonian dynamics and that the
tools of information geometry can be used to provide natural justifications
for both the metric of configuration space and for the particular form of the
quantum potential. In this paper these improvements --- the elimination of
auxiliary variables, the derivation of Hamiltonian dynamics, and the
introduction of information geometry methods --- are extended to the ED of
quantum scalar fields.

\section{Entropic Dynamics}

We wish to study the quantum dynamics of a single scalar field. A field
configuration $\phi(x)$ associates one real degree of freedom to each spatial
point $x$ in three-dimensional Euclidean space. Such a configuration is
represented as a point $\phi\in\mathcal{C}$ in an $\infty$-dimensional
configuration space $\mathcal{C}$ and it is convenient to represent the point
$\phi$ as a vector with infinitely many components denoted $\phi_{x}=\phi
(x)$.\footnote{$\infty$-infinite dimensional spaces are complicated objects.
We make no claim of mathematical rigor and follow the standard assumptions,
notation, and practices of the subject. To be definite we can, for example,
assume that the fields are initially defined on a discrete lattice (which
makes the dimension of $\mathcal{C}$ infinite but countable) and that the
continuum is eventually reached in the limit of vanishing lattice constant.}
In the ED framework the field $\phi_{x}$ has definite values which indicates a
major departure from the standard Copenhagen interpretation. However, in
general, the values $\phi_{x}$ are unknown and the objective is to determine
how their probability distribution $\rho\lbrack\phi]$ changes over time. The
first goal will be to use the method of maximum entropy to find the
probability $P[\phi^{\prime}|\phi]$ that the field configuration makes a
transition from a configuration $\phi$ to a neighboring configuration
$\phi^{\prime}$.

\paragraph*{The prior}

We start from a prior for the transition probability distribution
$Q[\phi^{\prime}|\phi]$ that expresses extreme ignorance: before any
information is taken into account the knowledge of how the field changes at
one point $x$ tells us nothing about how it changes at other points
$x^{\prime}$. This state of ignorance is represented by a prior that is a
product over all space points,
\begin{equation}
Q[\phi^{\prime}|\phi]\sim%
{\textstyle\prod\limits_{x}}
Q(\phi_{x}^{\prime}|\phi_{x})~. \label{prior}%
\end{equation}
Furthermore, we assume that for every point $x$ knowledge about the initial
$\phi_{x}$ tells us nothing about the final $\phi_{x}^{\prime}$. This is
represented by $Q(\phi_{x}^{\prime}|\phi_{x})\sim$ constant. Since such
constants have no effect on entropy maximization we can set $Q[\phi^{\prime
}|\phi]=1$.

\paragraph*{The constraints}

The actual information about evolution is introduced through constraints. The
first piece of information is that the evolution of the fields is continuous.
This means that at first we need only consider a small change; later we will
consider how a large change is achieved as a result of many small changes. For
each $x$ the field will change by a small amount from $\phi_{x}$ to $\phi
_{x}^{\prime}=\phi_{x}+\Delta\phi_{x}$ and we impose that the expected squared
change is
\begin{equation}
\left\langle \Delta\phi_{x}^{2}\right\rangle =\int D\phi^{\prime}\,P\left[
\phi^{\prime}|\phi\right]  \,\left(  \Delta\phi_{x}\right)  ^{2}=\kappa_{x}~,
\label{1Constr}%
\end{equation}
where $%
{\textstyle\int}
D\phi$ denotes a functional integration over $\mathcal{C}$. This is an
infinite number of constraints; one for each point $x$. The constant
$\kappa_{x}$ is some small number and a continuous motion will be eventually
achieved by letting $\kappa_{x}\rightarrow0$. To reflect the translational
invariance of three-dimensional Euclidean space we will set $\kappa_{x}%
=\kappa$ independent of $x$.

The constraints (\ref{1Constr}) lead to an evolution that is completely
isotropic in $\mathcal{C}$. Directionality is introduced assuming the
existence of a \textquotedblleft potential\textquotedblright\ $\Lambda
=\Lambda\lbrack\phi]$ and imposing a constraint on the expected displacement
$\left\langle \Delta\phi\right\rangle $ along the functional gradient of
$\Lambda$,
\begin{equation}
\left\langle \Delta\phi\right\rangle \cdot\nabla\Lambda\left[  \phi\right]
\equiv\int D\phi^{\prime}\,P\left[  \phi^{\prime}|\phi\right]  \,\int
d^{3}x\,\Delta\phi_{x}\frac{\delta\Lambda}{\delta\phi_{x}}=\kappa^{\prime}~,
\label{2Constr}%
\end{equation}
where $\delta/\delta\phi_{x}$ denotes the functional derivative and
$\kappa^{\prime}$ is a constant independent of $\phi$.

\paragraph*{Entropy maximization}

We seek the transition probability distribution $P\left[  \phi^{\prime}%
|\phi\right]  $ that maximizes the relative entropy%
\begin{equation}
S\left[  P,Q\right]  =-\int D\phi^{\prime}\,P\left[  \phi^{\prime}%
|\phi\right]  \log\frac{P\left[  \phi^{\prime}|\phi\right]  }{Q\left[
\phi^{\prime}|\phi\right]  }~ \label{Entropy}%
\end{equation}
\qquad subject to the constraints (\ref{1Constr}), (\ref{2Constr}), and
normalization. For $Q[\phi^{\prime}|\phi]=1$ the resulting distribution is Gaussian,%

\[
P\left[  \phi^{\prime}|\phi\right]  =\frac{1}{\zeta}\exp\left[  -\int
d^{3}x\left(  \frac{\alpha_{x}}{2}\left(  \Delta\phi_{x}\right)  ^{2}%
-\alpha^{\prime}\frac{\delta\Lambda}{\delta\phi_{x}}\Delta\phi_{x}\right)
\right]  ~,
\]
where $\alpha_{x}$ and $\alpha^{\prime}$ are Lagrange multipliers, and $\zeta$
is a normalization constraint. Since by translation invariance we had
$\kappa_{x}=\kappa$, the corresponding multipliers $\alpha_{x}$ must also be
independent of $x$ so that $\alpha_{x}=\alpha$. Furthermore, since both the
potential $\Lambda$ and the constant $\kappa^{\prime}$ are so far unspecified
we can, without loss of generality, absorb $\alpha^{\prime}$ into $\Lambda$
which amounts to setting $\alpha^{\prime}=1$. The resulting transition
probability is
\begin{equation}
P\left[  \phi^{\prime}|\phi\right]  =\frac{1}{Z}\exp\left[  -\frac{\alpha}%
{2}\int d^{3}x\left(  \Delta\phi_{x}-\frac{1}{\alpha}\frac{\delta\Lambda
}{\delta\phi_{x}}\right)  ^{2}\right]  \label{trans prob}%
\end{equation}
where $Z$ is a new normalization constant. In eq.(\ref{trans prob}) we see
that $\kappa\rightarrow0$ is recovered as $\alpha\rightarrow\infty$.

\paragraph*{Drift and fluctuations}

The transition probability (\ref{trans prob}) shows that a small change
$\Delta\phi_{x}$ can be written as an expected drift plus a fluctuation,
$\Delta\phi_{x}=\left\langle \Delta\phi_{x}\right\rangle +\Delta w_{x}$. The
expected drift is given by%

\begin{equation}
\left\langle \Delta\phi_{x}\right\rangle =\int D\phi^{\prime}\Delta\phi
_{x}P\left[  \phi^{\prime}|\phi\right]  =\frac{1}{\alpha}\frac{\delta\Lambda
}{\delta\phi_{x}}~. \label{drift}%
\end{equation}
The expected fluctuations are such that
\begin{equation}
\left\langle \Delta w_{x^{\prime}}\right\rangle =0\quad\text{and}%
\quad\left\langle \Delta w_{x}\Delta w_{x^{\prime}}\right\rangle =\frac
{1}{\alpha}\delta_{xx^{\prime}}~, \label{fluct}%
\end{equation}
where $\delta_{xx^{\prime}}=\delta(x-x^{\prime})$. Since $\Delta w_{x}%
\sim\alpha^{-\frac{1}{2}}$ while $\left\langle \Delta\phi_{x}\right\rangle
\sim\alpha^{-1}$ we see that for large $\alpha$ the fluctuations dominate the dynamics.

\paragraph*{Entropic Time}

In ED time is introduced as a book-keeping device to keep track of the
accumulation of small changes. This involves introducing a notion of instants
that are distinct and ordered, and defining the interval or duration between
them. For details see \cite{Caticha 2010}\cite{Caticha 2010b}. The result is
that if $\rho_{t}[\phi]$ refers to a probability distribution at a given
instant, which we label $t$, then entropic time is constructed by defining the
\emph{next} instant, labelled $t^{\prime}$, in terms of a distribution
$\rho_{t^{\prime}}[\phi^{\prime}]$ given by
\begin{equation}
\rho_{t^{\prime}}\left[  \phi^{\prime}\right]  =\int D\phi\,P\left[
\phi^{\prime}|\phi\right]  \rho_{t}\left[  \phi\right]  \label{Inst2}%
\end{equation}
where $P\left[  \phi^{\prime}|\phi\right]  $ is given by (\ref{trans prob}).
This definition readily lends itself to an iterative process in which time is
constructed instant by instant: $\rho_{t^{\prime}}$ is constructed from
$\rho_{t}$, $\rho_{t^{\prime\prime}}$ is constructed from $\rho_{t^{\prime}}$,
and so on. This process defines the dynamics.

It remains to specify the interval $\Delta t$ between two successive instants
$t$ and $t^{\prime}$ and the idea is captured by Wheeler's slogan: \emph{time
is defined so that motion }(or, in our case, the evolution of the
fields)\emph{\ looks simple}. For small changes the dynamics is dominated by
the fluctuations, eq.(\ref{fluct}). It is therefore convenient to define
duration so that the fluctuations are simple. Let%
\begin{equation}
\alpha=\frac{1}{\eta\Delta t}\quad\text{so that}\quad\left\langle \Delta
w_{x}\Delta w_{x^{\prime}}\right\rangle =\eta\Delta t\,\delta_{xx^{\prime}}\,,
\label{alpha fluct}%
\end{equation}
where $\eta$ is a constant (which will eventually be regraduated into $\hbar$)
that fixes the units of time relative to those of $\phi$.
\ \ \ \ \ \ \ \ \ \ \ \ \ \ \ \ \ \ \ \ \ \ \ \ \ \ \ \ \ \ \ \ \ \ \ \ \ \ \ \ \ \ \ \ \ \ \ \ \ \ \ \ \ \ \ \ \ \ \ \ \ \ \ \ \ \ \ \ \ \ \ \ \ \ \ \ \ \ \ \ \ \ \ \ \ \ \ \ \ \ \ \ \ \ \ \ \ \ \ \ \ \ \ \ \ \ \ Thus,
just as in Newtonian mechanics time is defined so that a free particle travels
equal distances in equal times, in the ED of fields time is defined so that
the fields undergo equal fluctuations in equal times. The translation
invariance ($\alpha_{x}=\alpha$) guarantees that time flows at the same rate everywhere.

\paragraph*{The information geometry of configuration space}

To each point $\phi\in\mathcal{C}$ we can associate a probability distribution
$P[\phi^{\prime}|\phi]$. Therefore $\mathcal{C}$ is a statistical manifold and
up to an arbitrary global scale factor its geometry is uniquely determined by
the information metric,%
\begin{equation}
\gamma_{xx^{\prime}}=C\int D\phi^{\prime}\,P[\phi^{\prime}|\phi]\frac
{\delta\log P[\phi^{\prime}|\phi]}{\delta\phi_{x}}\frac{\delta\log
P[\phi^{\prime}|\phi]}{\delta\phi_{x^{\prime}}}~, \label{gamma C}%
\end{equation}
where $C$ is an arbitrary positive constant. (See \emph{e.g.},\cite{Caticha
2012}.) For short steps ($\alpha\rightarrow\infty$) a straightforward
substitution of (\ref{trans prob}) using (\ref{alpha fluct}) yields
\begin{equation}
\gamma_{xx^{\prime}}=\frac{C}{\eta\Delta t}\delta_{xx^{\prime}}~.
\label{gamma}%
\end{equation}
We see that as $\Delta t\rightarrow0$ we have $\gamma_{xx^{\prime}}%
\rightarrow\infty$. The reason is that as the distributions $P[\phi^{\prime
}|\phi]$ and $P[\phi^{\prime}|\phi+\Delta\phi]$ become more sharply peaked it
becomes increasingly easier to distinguish one from the other which means the
information distance between them diverges. To define a distance that remains
meaningful for arbitrarily small $\Delta t$ it is convenient to choose
$C=\eta\Delta t$. Thus the metric $\gamma_{xx^{\prime}}=\delta_{xx^{\prime}}$
of the configuration space $\mathcal{C}$ is a straightforward generalization
of the metric $\delta_{ij}$ of Euclidean space and the distance $\Delta\ell$
between two slightly different configurations $\phi$ and $\phi+\Delta\phi$ is
\begin{equation}
\Delta\ell^{2}=\int d^{3}xd^{3}x^{\prime}\,\delta_{xx^{\prime}}\Delta\phi
_{x}\Delta\phi_{x^{\prime}}=\int d^{3}x\,(\Delta\phi_{x})^{2}~.
\label{distance}%
\end{equation}
In \cite{Caticha 2012} this choice of distance was merely postulated; here it
is justified from information geometry, the assumptions implicit in
(\ref{prior}), (\ref{1Constr}), and translation invariance.

\paragraph*{The Fokker-Planck equation}

The dynamics expressed by the integral equation (\ref{Inst2}) can be rewritten
in differential form. The result is a functional Fokker-Planck equation (see
\emph{e.g.}, \cite{Caticha 2010b}) that takes the form of a continuity
equation,%
\begin{equation}
\partial_{t}\rho_{t}\left[  \phi\right]  =-\int d^{3}x\frac{\delta}{\delta
\phi_{x}}\left(  \rho_{t}\left[  \phi\right]  v_{x}\left[  \phi\right]
\right)  ~. \label{FP}%
\end{equation}
(The combination $\int d^{3}x\frac{\delta}{\delta\phi_{x}}$ is the functional
version of the divergence.) The velocity $v_{x}\left[  \phi\right]  $ with
which probabilities propagate in configuration space is called the current
velocity. It is given by
\begin{equation}
v_{x}\left[  \phi\right]  =b_{x}\left[  \phi\right]  +u_{x}\left[
\phi\right]  \ , \label{CurrVel1}%
\end{equation}
where%

\begin{equation}
b_{x}\left[  \phi\right]  =\frac{\left\langle \Delta\phi_{x}\right\rangle
}{\Delta t}=\eta\frac{\delta\Lambda}{\delta\phi_{x}}\quad\text{and}\quad
u_{x}\left[  \phi\right]  =-\eta\frac{\delta\log\rho^{1/2}}{\delta\phi_{x}}~,
\end{equation}
are the drift and the osmotic velocities. The current velocity $v_{x}\left[
\phi\right]  $ can be written as the functional gradient of a scalar
functional $\Phi$,%
\begin{equation}
v_{x}\left[  \phi\right]  =\frac{\delta\Phi}{\delta\phi_{x}}\text{\quad
where\quad}\frac{\Phi\lbrack\phi]}{\eta}=\Lambda\left[  \phi\right]  -\log
\rho^{1/2}\left[  \phi\right]  ~. \label{CurrVel2}%
\end{equation}

Incidentally, it is convenient to introduce a functional $H[\rho,\Phi]$ on
$\mathcal{C}$ in order to write the Fokker-Planck equation as a functional
derivative in $\mathcal{C}$,
\begin{equation}
\partial_{t}\rho\left[  \phi\right]  =\frac{\Delta H\left[  \rho,\Phi\right]
}{\Delta\Phi\lbrack\phi]}~. \label{Hamilton a}%
\end{equation}
(For a useful and brief description of functional calculus in configuration
space see \cite{Hall et al 2003}.) Using (\ref{FP}), equation
(\ref{Hamilton a}) is easily integrated. The result is%

\begin{equation}
H\left[  \rho,\Phi\right]  =\int D\phi\,\int d^{3}x\frac{1}{2}\rho\left(
\frac{\delta\Phi}{\delta\phi_{x}}\right)  ^{2}+F[\rho]~, \label{Hamiltonian}%
\end{equation}
where $F[\rho]$ is an integration constant. In what follows we will assume
that $F[\rho]$ is independent of time. We emphasize that eq.(\ref{Hamilton a})
does not reflect a new assumption or a new dynamical principle; it is merely a
rewriting of (\ref{FP}).

\section{Non-dissipative Diffusion}

The Fokker-Planck equation (\ref{FP}) describes a standard diffusion process,
it does not describe quantum systems. As discussed in \cite{Caticha
2010}\cite{Caticha et al 2014b} the solution to this problem is to modify the
constraints: instead of $\Lambda\lbrack\phi]$ being an externally prescribed
potential we allow it to represent a dynamical field on $\mathcal{C}$. The
appropriate constraint consists in demanding that at each instant of time the
potential $\Lambda$, or equivalently the related quantity $\Phi$ in
(\ref{CurrVel2}), is updated in such a way that a certain functional -- that
we will call \textquotedblleft energy\textquotedblright\ -- remains constant.
It turns out that the appropriate \textquotedblleft energy\textquotedblright%
\ is the functional $H[\rho,\Phi]$ given in eq.(\ref{Hamiltonian}). Thus, the
dynamics consists of the coupled non-dissipative evolution of $\rho\lbrack
\phi]$ and $\Phi\lbrack\phi]$.

\paragraph*{The ensemble Hamiltonian and its conservation}

To impose a non dissipative diffusion we demand the conservation of the
functional $H\left[  \rho,\Phi\right]  $,
\begin{equation}
\frac{dH\left[  \rho,\Phi\right]  }{dt}=\int D\phi\,\left[  \frac{\Delta
H}{\Delta\Phi}\partial_{t}\Phi+\frac{\Delta H}{\Delta\rho}\partial_{t}%
\rho\right]  =0~.
\end{equation}
Using (\ref{Hamilton a})
\begin{equation}
\frac{dH\left[  \rho,\Phi\right]  }{dt}=\int D\phi\,\left[  \partial_{t}%
\Phi+\frac{\Delta H}{\Delta\rho}\right]  \partial_{t}\rho=0~. \label{HCons}%
\end{equation}
This condition must be satisfied at all times $t$ and for arbitrary choices of
the initial values of $\rho$ and $\Phi$. From (\ref{FP}) this means that
(\ref{HCons}) must hold for arbitrary choices of $\partial_{t}\rho$ which
implies that the integrand of (\ref{HCons}) must vanish. Therefore,
\begin{equation}
\partial_{t}\Phi=-\frac{\Delta H}{\Delta\rho}\text{ \ \ \ and \ \ }%
\partial_{t}\rho=\frac{\Delta H}{\Delta\Phi}~, \label{Ham eqs}%
\end{equation}
which we recognize as a functional form of Hamilton's equations with the
conserved functional $H[\rho,\Phi]$ playing the role of Hamiltonian.

\paragraph*{The Schr\"{o}dinger functional equation}

The Fokker-Planck equation together with the conservation of $H\left[
\rho,\Phi\right]  $ leads to a Hamiltonian structure regardless of the choice
of $F[\rho]$. However, as discussed in \cite{Caticha et al 2014b}, quantum
theory is reproduced only for a special choice of $F[\rho]$,
\begin{equation}
F[\rho]=\int D\phi\,\int d^{3}x\left[  \rho V\left(  \phi_{x},\nabla\phi
_{x}\right)  +\frac{\xi}{\rho}\left(  \frac{\delta\rho}{\delta\phi_{x}%
}\right)  ^{2}\right]  ~. \label{F[rho]}%
\end{equation}
In the first term $V\left(  \phi_{x},\nabla\phi_{x}\right)  $ is a potential
energy density to be discussed further below. The second term is usually
called the \textquotedblleft quantum\textquotedblright\ potential. It is the
functional trace of the Fisher information and its origin in information
geometry is discussed in \cite{Caticha et al 2014b}. $\xi$ is a positive
constant that controls the effect of the quantum potential.

As a matter of convenience we can combine the two variables $\rho\lbrack\phi]$
and $\Phi\lbrack\phi]$ into a single complex variable, $\Psi_{k}[\phi
]=\rho^{1/2}e^{ik\Phi/\eta}$, where $k$ is an arbitrary positive constant. The
pair of Hamilton's equations (\ref{Ham eqs}) can then be combined into a
single non-linear equation for the wave functional $\Psi_{k}\left[
\phi\right]  $,%

\[
i\frac{\eta}{k}\partial_{t}\Psi_{k}\left[  \phi\right]  =\int d^{3}x\left[
-\frac{\eta^{2}}{2k^{2}}\frac{\delta^{2}}{\delta\phi_{x}^{2}}+\left(
\frac{\eta^{2}}{2k^{2}}-4\xi\right)  \frac{1}{|\Psi_{k}|}\frac{\delta^{2}%
|\Psi_{k}|}{\delta\phi_{x}^{2}}+V\right]  \Psi_{k}\left[  \phi\right]  ~.
\]
Different choices of the arbitrary $k$ lead to different but equivalent
descriptions of the same theory. Let us therefore take advantage of the
arbitrariness of $k$ and choose the simplest and most convenient description.
This is achieved for the value $\hat{k}=(\eta^{2}/8\xi)^{1/2}$ that leads to
the linear Schr\"{o}dinger equation,
\begin{equation}
i\hbar\partial_{t}\Psi\left[  \phi\right]  =\int d^{3}x\left[  -\frac
{\hbar^{2}}{2}\frac{\delta^{2}}{\delta\phi_{x}^{2}}+V\right]  \Psi\left[
\phi\right]  ~,
\end{equation}
where we have identified $\eta/\hat{k}=\hbar$ and dropped the index $k$ so
that $\Psi=\rho^{1/2}e^{i\Phi/\hbar}$. This is quantum field theory in the
Schr\"{o}dinger representation and one can now proceed in the usual way to
introduce a Hilbert space, operators, and all the standard machinery of
quantum mechanics. For example, the commutator of the field $\phi_{x}$ and its
conjugate momentum is
\[
\lbrack\phi_{x},\frac{\hbar}{i}\frac{\delta}{\delta\phi_{x^{\prime}}}%
]=i\hbar\delta_{xx^{\prime}}~.
\]

At this point the potential $V(\phi_{x},\nabla\phi_{x})$ is essentially
arbitrary. A useful form is obtained by doing a Taylor expansion about weak
fields and gradients and then imposing the rotational and Lorentz symmetries
required by the experimental evidence,%

\begin{equation}
V(\phi_{x},\nabla\phi_{x})=\frac{1}{2}(\nabla\phi_{x})^{2}+\frac{1}{2}%
m^{2}\phi_{x}^{2}+\lambda^{\prime}\phi_{x}^{3}+\lambda^{\prime\prime}\phi
_{x}^{4}+\ldots\label{potential}%
\end{equation}
The various coefficients represent mass and other coupling constants. We
conclude that the ED framework reproduces the Schr\"{o}dinger representation
of the standard relativistic quantum theory of scalar fields.\cite{Jackiw
1989}

\section{Discussion}

Setting $\lambda^{\prime}=\lambda^{\prime\prime}=\ldots=0$ the Schr\"{o}dinger
equation,
\begin{equation}
i\hbar\partial_{t}\Psi=\frac{1}{2}\int d^{3}x\left[  -\hbar^{2}\frac
{\delta^{2}}{\delta\phi_{x}^{2}}+\left(  \partial\phi_{x}\right)  ^{2}%
+m^{2}\phi_{x}^{2}\right]  \Psi~,
\end{equation}
reproduces the quantum theory of free real scalar fields \cite{Jackiw 1989}
and all the standard results can now be obtained using conventional methods
(see \emph{e.g.}, \cite{Long Shore 1998}). For example, choosing units such
that $\hbar=c=1$, a standard calculation of the ground state gives a Gaussian
functional,%
\begin{equation}
\Psi_{0}\left[  \phi\right]  =\frac{1}{Z_{0}^{1/2}}e^{-iE_{0}t}\exp\left[
-\frac{1}{2}\int d^{3}x\int d^{3}y\,\,\phi\left(  \vec{x}\right)  G\left(
\vec{x},\vec{y}\right)  \phi\left(  \vec{y}\right)  \right]  ~,
\end{equation}
where
\begin{equation}
G(\vec{x},\vec{y})=\int\frac{d^{3}k}{(2\pi)^{3}}\omega_{k}\,e^{i\vec{k}%
\cdot(\vec{x}-\vec{y})}~,\quad\text{with\quad}\omega_{k}=(\vec{k}^{2}%
+m^{2})^{1/2}~.
\end{equation}
The energy of the ground state is%

\begin{equation}
E_{0}=\left\langle H\right\rangle _{0}=\frac{1}{2}\int d^{3}x\,G\left(
\vec{x},\vec{x}\right)  =\int d^{3}x\int\frac{d^{3}k}{\left(  2\pi\right)
^{3}}\frac{1}{2}\omega_{k}%
\end{equation}
is both infrared and ultraviolet divergent. The vacuum expectation value of
the field at any point $\vec{x}$ vanishes while its variance diverges,%
\begin{equation}
\left\langle \phi\left(  \vec{x}\right)  \right\rangle =0\quad\text{and}%
\quad\text{Var}\left[  \phi\left(  \vec{x}\right)  \right]  =\langle\phi
^{2}\left(  \vec{x}\right)  \rangle_{0}=\int\frac{d^{3}k}{\left(  2\pi\right)
^{3}}\frac{1}{2\omega_{k}}~.
\end{equation}
Note, however, that what diverges here are not the physical fields but the
uncertainty in our predictions. ED recognizes the role of incomplete
information: the theory is completely unable to predict the field value at a
sharply localized point. The theory does, however, offer meaningful
predictions for other quantities. For example, the equal time correlations
between two field variables $\phi\left(  \vec{x}\right)  $ and $\phi\left(
\vec{y}\right)  $ are \cite{Long Shore 1998},%
\begin{equation}
\left\langle \phi\left(  \vec{x}\right)  \phi\left(  \vec{y}\right)
\right\rangle _{0}=\int\frac{d^{3}k}{\left(  2\pi\right)  ^{3}}\frac
{e^{i\vec{k}\cdot\left(  \vec{x}-\vec{y}\right)  }}{2\omega_{k}}=\frac{m}%
{4\pi^{2}\left\vert \vec{x}-\vec{y}\right\vert }K_{1}\left(  m\left\vert
\vec{x}-\vec{y}\right\vert \right)
\end{equation}
where $K_{1}$ is a modified Bessel function.

\paragraph*{Conclusion}

Entropic dynamics provides an alternative method of quantization --- entropic
quantization. In the ED framework a quantum theory is a non-dissipative
diffusion in the configuration space.

The entropic quantization of scalar fields yields the standard predictions of
quantum field theory. At this early point in the development the advantages of
the entropic approach do not lie in any new predictions (at least not yet) but
rather in the suitability of the formalism to be extended beyond the domain in
which ED is equivalent to the current quantum field theory and in the new
insights it offers on matters of interpretation. More specifically, concerning
entropic time: In the ED of fields, the field fluctuations provide the clock
and entropic time is defined so that field fluctuations are uniform in space
and time. Concerning the nature of particles: fields are real, particles are
just some peculiar spatial correlations in the field.\ Concerning the
divergences: they are the expected consequence of handling incomplete
information. Some predictions will be certain, some will be uncertain, and
some may even be infinitely uncertain.

\paragraph*{Acknowledgments}

We would like to thank D. Bartolomeo, C. Cafaro, N. Caticha, S. DiFranzo, A.
Giffin, P. Goyal, D.T. Johnson, K. Knuth, S. Nawaz, M. Reginatto, C.
Rodr\'{\i}guez, and J. Skilling for many discussions on entropy, inference and
quantum mechanics.

\end{document}